# Mapping the free energy of lithium solvation in the protic ionic liquid Ethylammonuim Nitrate: A metadynamics study

Ali Kachmar*[a], Marcelo Carignano[a], Teodoro Laino[b], Marcella Iannuzzi[c], Jürg Hutter[c]

[a]Qatar Environment and Energy Research Institute, Hamad Bin Khalifa Universtiy, Qatar Foundation, Doha, Qatar

[b]Industry Solutions and Cognitive Computing, IBM Zurich Research Laboratory and [c]Department of Chemistry, University of Zurich, Winterthurerstrasse 190, CH-8057, Zurich, Switzerland

Email: akachmar@qf.org.qa, akachmar@hbku.edu.qa

## Abstract

The understanding of lithium solvation and transport in ionic liquids is important due to the possible applications in electrochemical devices. Using first principles simulations aided with the metadynamics approach we study the free energy landscape for lithium at infinite dilution conditions in ethylammonium nitrate, a protic ionic liquid. We analyze the local structure of the liquid around the lithium cation and find a quantitative picture in agreement with experimental findings. Our simulations show that the lowest two free energy minima correspond to conformations with the lithium solvated either by 3 or 4 nitrates ions with a transition barrier between them of 0.2 eV. Other less probable conformations having a different solvation pattern are also investigated.

# I. Introduction

Ionic liquids (ILs) have attracted the interest of the academic community for many years. However, the possibility of designing ILs for specific applications is now being considered from many different technological areas. One example is the Li-ion battery (LIB) industry that needs to address safety issues emerging from the flammability of the currently used carbonate based organic electrolytes. Indeed, some ILs are good candidates as safer electrolytes because they are non-flammable. Moreover, ILs in general have negligible vapor pressure and high chemical and thermal stability, properties that are also beneficial for the reliability of LIBs. The essential role of the electrolyte is to conduct the electric charges between the electrodes by allowing the transport of ions. In that respect, ILs offer a concrete possibility for LIBs and other electrochemical storage devices [1-3]. Understanding the ion transport mechanisms is of great importance not only for LIBs and fuel cells but also for many other systems [4-6]. Interfaces involving ILs [7-9] and computational design of new electrolyte materials [10-12] are also a subject of great interest for the electrochemistry community. Protic ionic liquids (PILs) are a subset of ionic liquids that have recently gained widespread scientific attention [13-16]. Balducci et al. suggested that PILs represent a new and interesting class of electrolytes for LIBs [17-19] because of their large conductivity with respect to the more common aprotic ones. Moreover, PIL are in general less expensive than their aprotic counterparts, they are relatively easy to synthesize and environmentally friendly. All these properties are attractive for the modern industry and in particular for LIB applications [20].

In this work we present our study on the Li-ion solvation and diffusion in EthylAmmonium Nitrate (EAN) by means of ab initio molecular dynamics (AIMD) based on density functional theory (DFT). EAN is a structurally heterogeneous and bicontinuous PIL, with polar domains, constituted of nitrate anion and ammonium groups, and apolar domains, constituted of methylene and methyl groups. The packing into an ordered liquid is provided by the solvophobic, electrostatic and strong hydrogen bonding interactions [21,22] as for other pure ILs [23-38]. The melting point is at 286 K [39,40].

The transport of ions in the electrolyte is one key phenomenon governing the performance Li-ion batteries. Nevertheless, the understanding of lithium solvation and its transport in PIL at the atomistic level is still lacking. Previous AIMD simulations [41,42] addressed the dynamical properties of neat EAN, revealing the presence of ions rattling in Long Living (LL) cages, which are formed by stable LL-pairs. The lithium solvation problem, instead, has only been studied with MD simulations based on Empirical Force Fields (EFF) [43-46]. Moreover, the characterization of the Li solvation environment has been attempted by means of neutron diffraction experiments [44]. In spite of all these valuable efforts, the complexity of the Li solvation and its mobility through the viscous EAN structure has not been fully disentangled yet.

There have been some interesting studies on the Li+ solvation and its transport properties in liquids, liquid electrolytes, and different kind of ionic liquids [48-58]. There were also some interesting studies explicitly on the structure and the dynamics of molten lithium nitrate ($LiNO_3$). The first one is by Kato et al, who discussed the rotational, translational and re-orientational motions of molten $LiNO_3$ using classical potentials [59-61]. Later, Adya et al investigated the structure of the $LiNO_3$ combining MD simulations with X-ray and neutron diffraction experiments. They reported that the lithium monovalent cation is surrounded on an average by four nitrate ($NO_3^-$) ions, with one oxygen atom in each of these $NO_3^-$ ions facing towards the $Li^+$ cation [62]. Kameda et al, arrived to a similar conclusion and proposed that the lithium is also surrounded tetrahedrally by four nitrate ions, and one oxygen atom in each of these four ($NO_3^-$) ions points towards the $Li^+$ cation [63].

In this study we use AIMD in combination with the metadynamics (MTD) [64-67] approach in order to shed light on the solvation structure and diffusion pathways for lithium transport in EAN [68-72]. The simulations are sufficiently long in order to sample the different solvation possibilities. We show that the most favorable lithium scenarios are associated with a coordination with 3 and 4 nitrates, which in turn, are associated to configurations having a polulation of monodentate and bidentate geometries. We also discuss other lithium local environments through the analysis of the radial distribution functions and the free energy landscape along with the minimum energy pathway connecting the two most probable local minima.

The structure of the paper is as follows. The computational details and the description of the employed model system are in Section II. The structural characterization of the generated trajectory and the discussion of the reconstructed free energy surface (FES) are reported in Section III.

## II. Computational details and model systems

This study relies mainly on AIMD simulations carried out with the CP2K [73-82] package. However, the preparation of the model has been carried out at the EFF level of theory, using the Gromacs package [83-86]. The chosen parameterization is the one proposed by Varela et al. [43]. The EFF-MD simulations are meant to generate a starting structure for AIMD without excessively consuming computational resources.

The model system contains 39 EAN ion pairs plus one $LiNO_3$, (590 atoms) and it is globally electrically neutral. Hence, it is the equivalent of ~ 1.61 wt % solution $LiNO_3$ in EAN with a density of 1206.2313 kg/m$^3$. The available experiments were performed on a ~ 6 wt % solution $LiNO_3$ in EAN [47].

The ab initio MD simulations are carried out within the Born-Oppenheimer approximation. The electronic structure properties were calculated using the PBE[87] functional with the empirical correction from Grimme scheme (DFT-D3)[88,89] to account for the dispersion interactions, which are important in ionic liquids [90-92]. Kohn-Sham orbitals are expanded in a Gaussian basis set for the all atoms (MOLOPT-DZVP-SR-GTH for Li, C, N, O, H), and we employed the norm-conserving GTH pseudopotentials. The auxiliary plane wave (PW) basis set was defined by the energy cutoff of 350 Ry. The equations of motion are integrated via the velocity Verlet algorithm with a time step of 1 fs. The temperature is controlled by the Nosé-Hoover chain of thermostats [93,94]. The simulations were extended up to 550 ps. Low spin states are used in all the calculations. The Brillouin zone was sampled at the $\Gamma$ point only throughout the present study. The system was firstly equilibrated with at least 10 ns at 298 using EFF-MD simulations, followed by a second equilibration which is started by sampling the constant pressure/constant temperature ensemble (NPT) at 300K for 20 ps where the pressure is controlled by means of the Parrinello-Rahman method [95] using the structure of liquid that was obtained from the previous EFF-MD equilibration. During this time, the simulation box slowly changes its size, approaching the experimental density of 1206.235 kg/m$^3$ after 10 ps [43]. After the NPT equilibration, we observe that the solvation shell of the Li$^+$ cation contains only 3 nitrates in a planar arrangement. This is actually at odds with the average coordination number of 4 recently reported in an EFF-MD study [43]. The general issue of the discrepancy between EFF and DFT simulations with respect to the solvation shell of ions in water has been extensively treated in the literature [96-97]. The deviation is often attributed to some polarization effect, which cannot be captured by the EFF description [98,99].

Starting from one snapshot at the experimental density extracted from the NPT trajectory, the sampling of the constant volume ensemble (NVT) is first carried out at a higher temperature (1500 K) for 6ps. From the high-T trajectory four configurations were selected corresponding to t=3 ps, 4 ps, 5 ps, and 6 ps. These have been used as initial coordinates for four independent annealing procedures, bringing the system from 1500 K to 300 K over 6 ps. After analyzing the four final local environments around the Li$^+$ cation, we selected as starting configuration for the further study one structure with four nitrates within the Li-solvation shell.

The production trajectory has been generated by running 36 ps in the NVT ensemble at 300 K, followed by more than 500 ps by applying the MTD acceleration scheme. MTD is used to investigate the rearrangements of the lithium coordination environment and the possible diffusion mechanism. This is done by defining a proper set of collective variables (CVs). These have to involve all the slow degrees of freedom that need to be activated in order to observe the relevant structural rearrangements under investigation. In our case, the CVs have to describe the local structure around the Li$^+$ cation. To this purpose, we selected two CVs. The first, $CV_1$, is the coordination between the Li and the oxygen atoms of the nitrate ions. The second one is the coordination between the Li$^+$ and the nitrogen atoms belonging to the nitrate ions ($CV_2$). These two CVs should provide the characterization of the solvation environment in terms of the number and relative orientation of the nitrate ions. The sampling in the reduced space of the CVs is enhanced by constructing the time dependent MTD potential. This potential is generated by spawning Gaussian hills along the trajectory. The height and the width of the Gaussians are 10$^{-3}$ and 0.2 in a.u, respectively. The time interval between spawning two consecutive hills equals 25 MD step.

# III. Results and discussion

## A. Lithium solvation and structure analysis

It is well known that pure ILs are nanostructured into polar and apolar domains [23-27]. Moreover, when other inorganic salts, like $Mg(NO_3)_2$, $Ca(NO_3)_2$, and $Al(NO_3)_2$, are dissolved into PILs, the metal cations ($Mg^{2+}$, $Ca^{2+}$, and $Al^{3+}$) are more likely to reside in polar domains [14,15,72,100,101]. This arrangement is also favored by the $Li^+$, which is found close to the polar groups and is strongly associated with the nitrate ions [43,47]. Also in EAN, the Li ion is known to be solvated by a first shell of nitrates, which screen it from the rest of the liquid. This first solvation shell is rather fluxional, i.e., the nitrates rotate and rearrange easily around the $Li^+$, and can also change in number or get exchanged.

The MD run is carried on in the NVT ensemble for the first 36 ps. During this time, the system explores the initial equilibrium state, where the coordination of Li to N spontaneously fluctuates between 3 and 4. The longest part of the trajectory, until the simulation time of 550 ps, is instead biased by the MTD procedure. The goal is to force larger fluctuations in the coordination shell of the $Li^+$, in order to investigate structural changes in its environment, like rotations and swaps of the nitrates, the hopping of the cation to other coordination pockets, prompting the diffusion mechanism. From the monitoring of the CVs (see Figure 1), we are able to identify the most relevant changes in the microstructure around the cation. The detailed molecular movements responsible for the CVs variations are revealed by closely inspecting the all-atom trajectory [102].

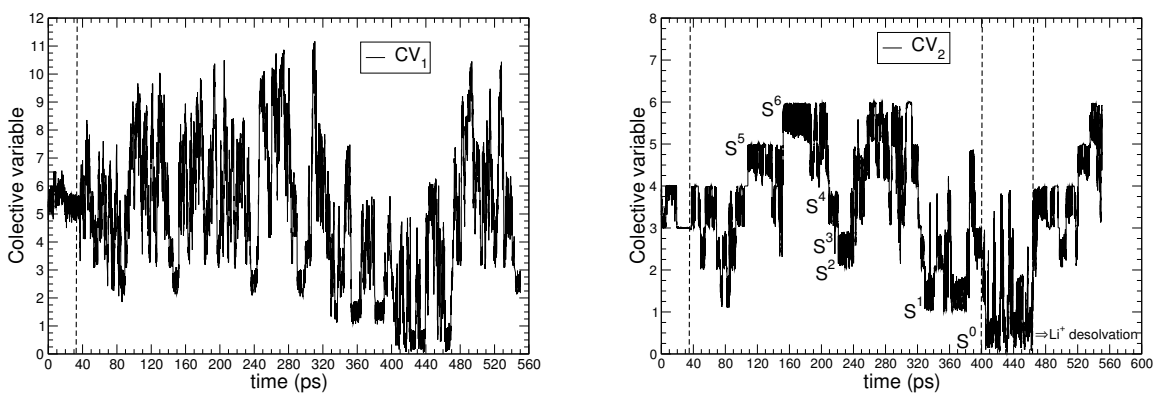

Figure. 1 Collective variables $CV_1$ (left hand side), and $CV_2$ (right hand side) in black, respectively, along the simulations trajectory. The dashed vertical line divides the unbiased and biased sections of the simulation. The states visited by the lithium during the MTD sampling are labeled as $S^0$, $S^1$, $S^2$, $S^3$, $S^4$, $S^5$, and $S^6$.

Along the NVT-MD, the number of nitrate molecules in the first coordination shell of the Li ion fluctuates between 3 and 4. The Li-O distances within the first coordination shell vary between 2.1 Å and 3.3 Å. All the fluctuations observed within the first 36 ps occur smoothly, without crossing significant barrier. The nitrate ion typically coordinates to the Li cation via the oxygen atoms, either in a monodentate geometry, i.e., only one O atom per $NO_3^-$ points to Li, or in a bidentate geometry, with two O atoms pointing to Li. In the monodendate configuration the Li-O distance is around 2.1 Å. In the bidendate configuration, the shortest distance is around 2.6 Å, while the second O is found between 3.1 and 3.3 Å. The same Li-N coordination can correspond to different local environments. For example, four nitrates in the first shell might give a Li-O coordination of 7, with three bidentate and one monodentate, or of 6, with two bidentate and 2 monodentate.

As a general remark, we observe that over the entire trajectory, the Li-O coordination changes more rapidly and shows much wider fluctuations. In particular, for the same number of nitrates in the first solvation shell, the Li-O coordination can vary when the local geometry changes from monodentate to bidentate. For example, four nitrates in the first shell might give a Li-O coordination of 7, with three bidendate and one monodendate, or of 6, with two bidendate and 2 monodendate.

The ion pairs lifetime in the EAN, has been reported to be more than 100 ps [41]. Therefore, in order to be able to observe processes that involve a broader reorganization within the liquid, with exchanges among the ion pairs, it is required to significantly extend the sampling. This can be achieved by applying the MTD formalism, as described above. Once the MTD is switched on, the system explores, indeed, a much broader area of the phase space. The large fluctuations in the Li-O coordination number are associated to the rotational and translational motion of the nitrate ions. Note that, once the MTD is activated, the range of variation becomes much wider, meaning that also the exchange between monodendate and bidendate geometries is enhanced. The large fluctuations of the two selected CVs suggest that the trajectory visits configurations corresponding to different solvation scenarios. From the detailed analysis of the microstructure's changes and the reconstruction of the underlying FES we are able to identify the transition region and possible new stable states.

The Li-N coordination changes more slowly and it can be used to distinguish among separate metastable states in the phase space. We label the identified states as follows: $S^0$(0 nitrate), $S^1$(1 nitrate), $S^2$(2 nitrates), $S^3$(3 nitrates), $S^4$(4 nitrates), $S^5$(5 nitrates), and $S^6$(6 nitrates). In all cases the fluctuations of the Li-N coordination are around +/- 0.3. Seven snapshots extracted from the MTD trajectory for each different coordination shell are illustrated in Figure 2, where we show only the Li-O distances in dashed lines. The $Li^+$-O average distances marked by the dashed lines are ~2.3 Å for the monodendate and 1.9 Å for the bidendate geometries. The corresponding $Li^+$-N distances are ~3.0 Å for the monodendate, and 2.3 Å for the bidendate geometries.

The Li-O coordination is often maximized, when the Li-N coordination is between 3 and 4. This means that the bidendate geometries are predominant. However, with larger Li-N coordination values, the Li-O coordination can become as small as 4. This situation corresponds to geometries where some nitrate ions within the solvation shell do not contribute to the oxygen coordination. This can be understood from pure steric effects: in order to have a larger number of nitrates closer to the $Li^+$, they rearrange such that the interactions with the oxygen atoms are disfavoured. On the other hand, large values in the oxygen coordination can also be associated to nitrates belonging to the second solation shell. These results imply that the presence of the lithium cation strongly affects the packing ordering in EAN.

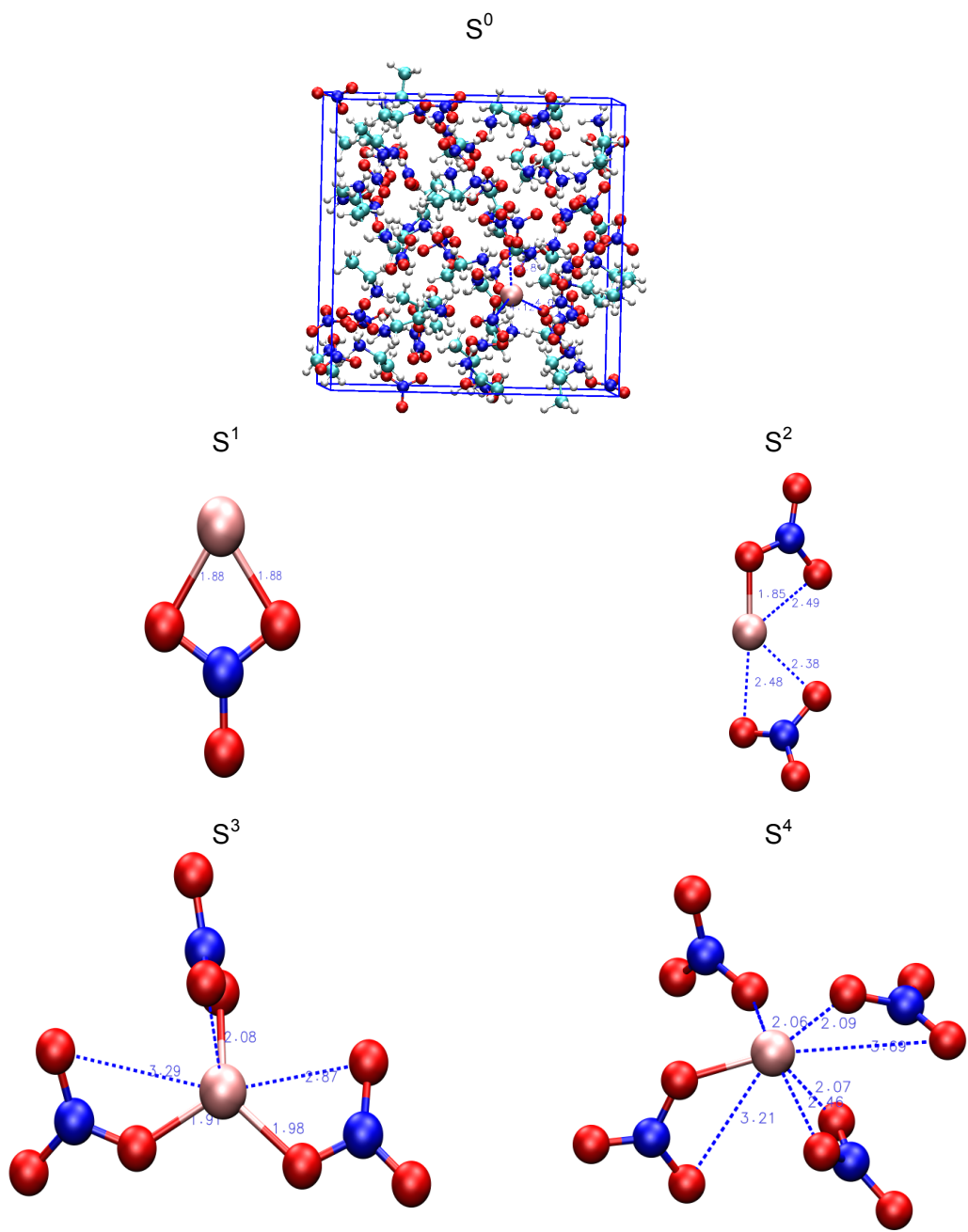

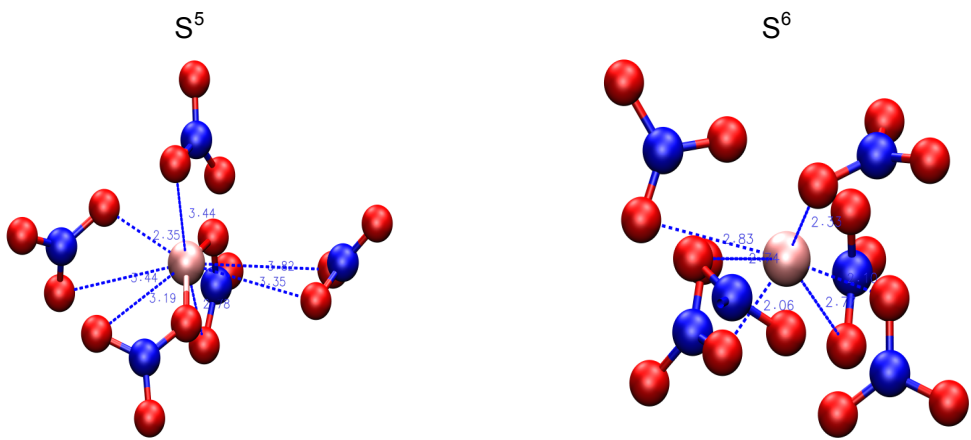

Figure. 2 Snapshots selected from the MTD trajectory showing a Li$^+$ cation solvated by 0 (S$^0$), 1(S$^1$), 2(S$^2$), 3(S$^3$), 4(S$^4$), 5(S$^5$), and 6 (S$^6$) nitrate ions in. These snapshots are given within a 5 Å radius of the Li$^+$ cation. Atoms outside the solvation shell are omitted for clarity. Atom colors: Li (pink), O (red), and N (blue).

Within the first 320 ps, the Li$^+$ experiences different local environments, while the number of nitrates in its coordination shell varies between 2 and six, with clear preference for the coordinations 3 and 4. The structural changes are determined by the rearrangements of the closest nitrates surrounding the lithium. The coordination to six nitrates has been already discussed in the literature [103].

At about 320 ps, the system undergoes an abrupt change. The Li-N coordination drops below 2 and below, and higher coordinations are stably recovered only after 480 ps. The Li ion is found in a diffusive state, surrounded by only few EhtylAmmonium molecules. This can be interpreted as a lithium desolvation state allowing the cation to move through the apolar region, until it can be newly fully solvated by other nitrates. This implies that the diffusion of the Li ion occurs by hopping between neighboring polar domains and not by dragging its own coordination shell. This finding is consistent with the high viscosity of such liquids, which prevents the moves over long distances of bulky pseudo-particles (i.e., Li$^+$ solvation).

The different possible coordinations within the same domain are repeatedly observed, with relatively frequent transitions from one to the other, suggesting that FES basins is rather smooth, corrugated just by small barriers. Therefore the kinetics within the domain is fast.

## B. Lithium coordination with nitrates

In order to better characterize the local environment of the Li$^+$ ion solvated in EAN also beyond the first solvation shell, we computed the Li-N pair distribution functions. Since the MTD trajectory samples regions of the phase space corresponding to different intermediate structures, the Li-N, g_N(r), radial distribution functions have been calculated over configurations belonging to a specific coordination environment. In particular, we distinguish four different situations, corresponding to the lithium coordination including 3, 4, 5, or 6 nitrates. The resulting radial distributions are plotted in Figure 3. When either 3 or 4 nitrates form the solvation shell, the g_N(r) shows a bimodal distribution, which is consistent with the easy switching between the bidendate and monodendate binding geometries. With coordination 4, two nitrates are always in the bidentate mode, one is always monodentate, and one switches between the two. Hence, first peak of the g_N(r) is slightly higher. For the 3-nitrates coordination, instead, the second peak is more pronounced, because the monodentate geometry is preferred. Hence, from the radial distribution function we could determine the most probable structure of the solvation shell.

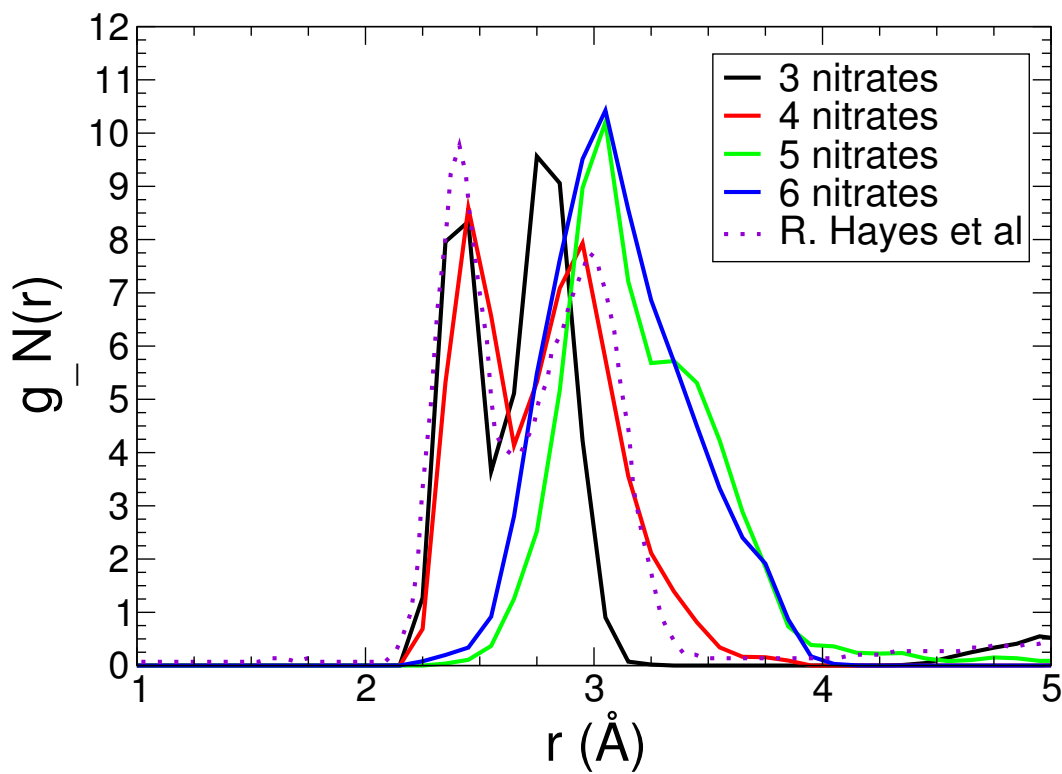

Figure. 3 Comparison of the radial distribution functions g_N(r) for the different lithium nitrate coordinations compared to the experimental data [47]

Analogously, when the lithium solvation shell is more populated, by 5 or 6 nitrates, the g_N(r) does not present the two peaks distribution, indicating that no molecule is in the bidendate geometry. This single peak is found at slightly longer distance, but still in the range of distances of the monodentate-binding mode.

R. Hayes et al. [47] considered that EA regions of the liquid are apolar, and the nitrates regions are polar. To characterize the polar environment of the lithium and its relation with the coordination number with nitrates, we calculated the integrated radial distribution function of g_$C_{CH3}...C_{CH3}$ for 0 =< $r_{max}$(Å) <= 5. The result of the integration shows that there are more EA (2.81) in coordination 4 than in coordination 3 (2.59). Then, the coordination 3 is characteried by a more polar environment than that of coordination 4.

## C. Free energy landscape for the lithium solvation in EAN

By performing metadynamics using a few CVs, it is possible to estimate the free energy surface (FES) in the reduced phase space defined by them. The FES is constructed progressively along the run, by exploring wider region of the phase space. When no more significant changes in the topography are observed, i.e., when no more new minima are discovered and the trajectory turns around among the same states, the run can be considered converged.

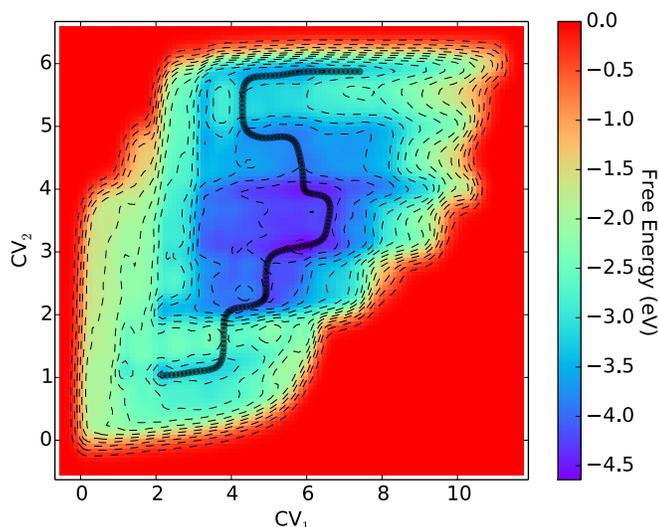

Figure. 4 Free energy surface as estimated from the MTD potential acting in the reduced phase space of the two selected collective variables. The contour lines are reported with a spacing in energy of 0.01 eV, the color map distinguish more probable areas (blue) from less visited areas (orange-red). The black dots mark the Minimum Energy Pathway (MEP) [104].

Our MTD run starts with a Li-N coordination between 3 and 4 and a Li-O coordination between 5 and 6. These states, indeed, appear as a large and deep well in the FES, in Figure. 4. The fine corrugation of this large basin of attraction, however, cannot be easily resolved. There are two well-defined local minima located in the basin, one more around $CV_2$ equal 3 and one at 4. The barrier separating these two states is less than 0.3 eV, and indeed could be crossed already at the standard MD level. As expected the available variability range of $CV_1$ is determined by the value of $CV_2$. As for the region with higher Li-N coordination number (e.g., ~6), the six surrounding nitrates can be either all monodendate or a combination between mono and bidentate configurations. The changes in coordination modes is very fluid and apparently barrier less. For this reason, in the two dimensional representation of the FES it is not possible to recognize any clear minimum. The region that shows the lowest Li-N coordination ($CV_2 < 1$) is associated with the lithium diffusion from the polar to the apolar domain. Also in this case, the 2D FES appears rather shallow, without clear structure, indicating a diffusive behavior. Indeed, in the time interval between 400 and 500 ps (see Fig.1), the fluctuations of both variables are continuous and wide between 0 and 3-5.

We show in Figure 5 the 1D MEP where we can distinguish the previously mentioned five states $S^2$, $S^3$, $S^4$, $S^5$, and $S^6$, of which only $S^2$, $S^3$, $S^4$, and $S^6$ can be identified as clear minima. The other state $S^5$ doesn't show a well-defined minimum rather transient configuration related to the high rotational mobility of the nitrates and to the diffusion of $Li^+$.

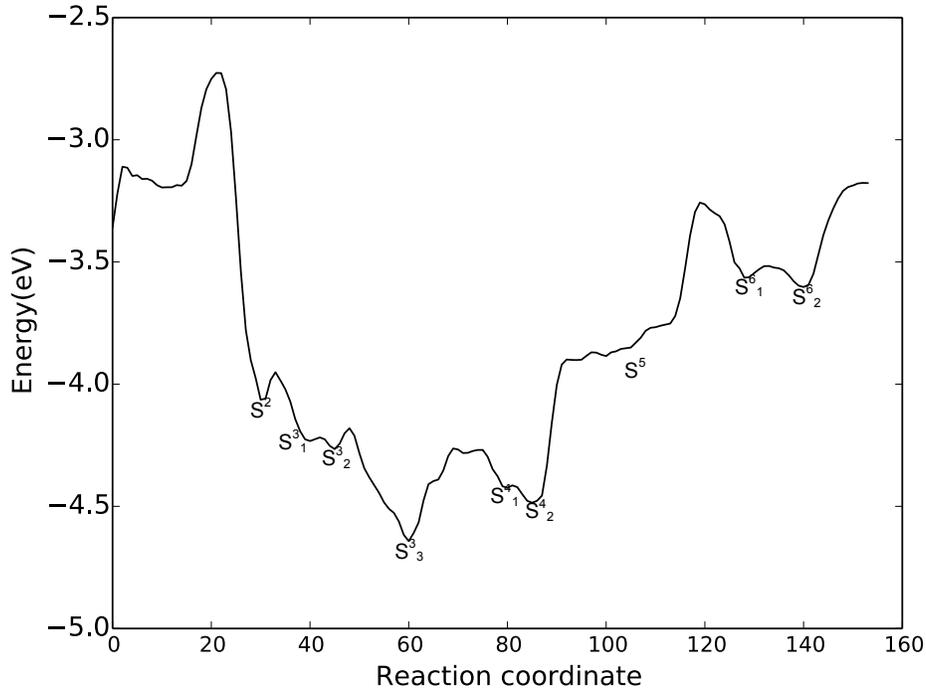

Figure. 5 One dimensional free energy surface

However, $S^3$, $S^4$, and $S^6$ show more than one minima which are labeled by $S^X_i$ where X=3, 4, or 6 and i could take 1, 2, and 3. The lithium coordination numbers (CN) for both oxygen ($CN^O$) and nitrates ($CN^N$) for all the minima found in the energy profile, are given in the table 1:

|  | $S^2$ | $S^3_1$ | $S^3_2$ | $S^3_3$ | $S^4_1$ | $S^4_2$ | $S^6_1$ | $S^6_2$ |
|---|---|---|---|---|---|---|---|---|
| $CN^N$ | 2 | 3 | 3 | 3 | 4 | 4 | 6 | 6 |
| $CN^O$ | 4 | 3.6 | 4 | 6 | 7 | 6 | 5 | 6 |
| Free energy (eV) | -4.06365 | -4.18685 | -4.21634 | -4.64245 | -4.42426 | -4.48538 | -3.56455 | -3.60253 |

The deepest energy minimum turns out to be the $S^3_3$, and it is associated to the lithium coordination with 3 nitrates in bidendate geometry. The transition between $S^3_3$ and $S^4_2$ crosses a free energy barrier of ~ 0.16 eV. However, the $S^4_1$, and $S^4_2$ are more stable than both $S^3_1$, and $S^3_2$. The fact that the $S^3$ and $S^4$ are showing more than one minima reflects the existence of a population of both mondendate and bidendate geometries, and the switching between them. Therefore, there is a clear competition between the lithium coordination with 3, and 4 nitrates respectively. It was previously reported [47] that every $Li^+$ cation is solvated by approximately four nitrates, and this is in line with our simulations.

The energy profile shows that higher coordination states are thermodynamically not stable. On the other hand, the $S^6$ has two minima $S^6_1$ and $S^6_2$, which are associated to the lithium coordination with 5, and 6 oxygens respectively. It is noticeable that in this coordination state $S^6$ the switching from the oxygen coordination 5 to 6 is barrier less. Yet, once in $S^6_{1,2}$ the lithium has to cross a barrier of about 0.88 eV to follow the pathway back into the most stable state $S^4_2$.

Finally, the MEP doesn't reveal any clear minimum for the lithium coordination 1, even though some structuring and corrugation can be observed. This metastable state is interesting because it shows the energy of the lithium de-solvation, necessary to allow the lithium diffusion in the liquid.

Relying on the findings of the biased and unbiased MD sampling and on what we observe in the MD/MTD trajectory, the lithium has to visit the solvation states located by the metadynamics while moving from a polar to a nearby apolar domain with a de-solvation energy computed to be around -3.33 eV, which is higher by about 1 eV than in the common organic electrolyte solvent [105]. This conclusion can be correlated to the fluctuation observed in the $CV_1$ (see Fig.1), which can be seen as a dragging force for the lithium diffusion in the liquid. Therefore, this is one of the possible mechanisms for lithium diffusion.

## IV. Conclusion

In this paper we report a first principles and metadynamics simulations study of the lithium solvation in EAN with an aim to understand the mechanisms involved in the solvation, and the diffusion of lithium. The main findings of our work are the molecular arrangement around the $Li^+$ cation, and the free energy landscape of the different lithium coordinations in this particular ionic liquid. Indeed, the classical MD suggested previously a solvation shell involving between four to five nitrate ions around a single $Li^+$. The performed unbiased molecular dynamics simulations in this study were not enough to have a good sampling of the lithium solvation environment due to the short time scale and the transient changes in the $CV_2$. However, the biased (MTD) simulations results leads to a different $Li^+$ solvation structure. The MTD simulations strongly show the predominant low free energy barriers for the lithium coordination with 3 nitrates ($S^3$), and 4 nitrates ($S^4$). The difference in free energy between these two states was found to be an activated process with a free energy barrier of ~ 0.2 eV. Moreover, the analysis of the g_N(r) for these two states implies a switching of the oxygen facing towards the $Li^+$ cation. The results of these analyses show also that the lithium is surrounded by an average of four or three nitrate ions in two different conformations corresponding to the nitrate ions coordinating with the lithium cation in either monodendate or bidendate geometries. In fact, the results of the structure and dynamics properties of the system discussed in this manuscript are also in good agreement with the recent experimental data [47]. Overall, these results are consistent with the findings from previous experiment for the same liquid but with higher lithium concentration.

This work represents a major computational effort that pushes the capabilities of the state-of-the-art AIMD to its limits. Extending studies like this one to many systems in a systematic way is still very expensive and therefore the use of simplified force fields is very important. In this sense, the current work provides a good standard to be used as a benchmark for the development of improved models.

# V. Acknowledgments

Ali Kachmar thanks QEERI for the laboratory abroad opportunity to visit the University of Zurich and IBM Zurich Research Laboratory, and also for Jesús Carrete for providing the MD trajectory and the force fields for the low lithium concentration configuration that was tested in the initial preparation of the model. Warm thanks are due to Sabre Kais, Ilias Belharouak, Mauro Boero, Michele Parrinello, Oleg Borodin and One-Sun Lee for stimulating discussions. The HPC resources and services used in this work were provided by the Research Computing group in Texas A&M University at Qatar. Research computing is funded by the Qatar Foundation for Education, Science and Community Development.

## VI. References

[1] P. Wasserscheid, A. Stark, Handbook of Green Chemistry, Volume **6**: Ionic Liquids. Copyright©(2010) WILEY-VCH Verlag GmbH & Co. KGaA, Weinheim, ISBN: 978-3-527-32592-4.

[2] A. Leawndowski, A. Świderska-Mocek, *J. Power. Sources*. **2009**, 194, 601-609.

[3] D. R. MacFarlane, N. Tachikawa, J. M. Pringle, P. C. Howlett, G. D. Elliott, J. H. Davis, M. Watanabe, P. Simon, C. Austen Angell, *Energy Environ. Sci.* **2014**, 7, 232-250.

[4] Y. Jiang, A. Lee, J. Chen, V. Ruta, M. Cadene, B. T. Chait, R. Mackinnon, Nature. **2003***, 423, 33-41.

[5] "The Nobel Prize in Chemistry 2003" (Press release). *The Royal Swedish Academy of Science*. 2003-10-08. Retrieved 2010-01-18.

[6] S. Ping Ong, L. J. Miara, J. Chul Kim, Y. Mo, G. Ceder, *Nature Materials*. **2015**, 14, 1026–1031.

[7] M. V. Fedorov, A. A. Kornyshev, A. *Chemical Reviews.* **2014**, 114(5), 2978-3036.

[8] B. Rotenberg, M. Salanne, *J. Phys. Chem. Lett.* **2015**, 6(24), 4978-4985.

[9] S. McDonald, A. Elbourne, G. G. Warr, R. Atkin, *Nanoscale*. **2016**, 8, 906-914.

[10] T. Husch, M. Korth, *Phys.Chem.Chem.Phys*., **2015**, 17, 22596.

[11] C. Schütter, T. Husch, V. Viswanathan, S. Passerini, A. Balducci, M. Korth, *Journal of Power Sources*, 326, **2016**, 541-548.

[12] X. Qu, A. Jain, N. N. Raiput, L. Cheng, Y. Zhang, S. P. Ong, M. Brafman, E. Maginn, L. A. Curtiss, K. A. Persson, *Computational Materials Science*, 2015, 103, 56–67.

[13] C. Austen Angell, N. Byrne, J.-P. Belieres, *Acc. Chem. Res.* **2007**, 40(11), 1228−1236.

[14] T. L. Greaves, C. J. Drummond, *Chem. Rev.* **2008**, 108, 206−237.

[15] T. L. Greaves, C. J. Drummond, *Chem. Rev.* **2015**, 115(20), 11379-11448.

[16] H. Matsuoka, H. Nakamoto, Md. Susan, M. Watanabe, *Electrochim. Acta*. **2005**, 50, 4015-4021.

[17] S. Menne, J. Pires, M. Anouti, A. Balducci, *Electrochemistry Communications*. **2013**, 31, 39-41.

[18] T. Vogi, S. Menne, R. S. Kuhnel, A. Balducci, *J. Mater. Chem. A*. **2015**, 8, 8258-8265.

[19] S. Menne, T. Vogl, A. Balducci, *Phys. Chem. Chem. Phys.* **2014**, 16, 5485-5489.

[20] A. Noda, Md. Susan, K. Kudo, S. Mitsushima, K. Hayamizu, M. Watanbe, J. *Phys. Chem. B*. **2003**, 107, 4024-4033.

[21] R. Atkin, G. G. Warr, *J. Phys. Chem. B*. **2008***, 112, 4164-4166.

[22] Y. Umebayashi, W.-L. Chung, T. Mitsugi, S. Fukuda, M. Takeuchi, K. Fujii, T. Takamuku, R. Kanzaki, S.-i. Ishiguro, *J. Comput. Chem. Jpn*. **2008***, 7, 125-134.

[23] Y. Wang, G. A. Voth, *J. Am. Chem. Soc*. **2005**, 127, 12192−12193.

[24] Y. Wang, G. A. Voth, *J. Phys. Chem. B*. **2006**, 110, 18601-18608.

[25] Y. Wang, W. Jiang, T. Yan, G. A. Voth, *Acc. Chem. Res*. **2007**, 40, 1193-1199.

[26] J. N. A. Canongia Lopes, A. A. H. Pádua, *J. Phys. Chem. B*. **2006***, 110, 3330−3335.

[27] K. Shimizu, M. F. Costa Gomes, A. A. H. Pádua, L. P. N. Rebelo, J. N.

Canongia Lopes, *J. Mol. Struct.: THEOCHEM.* **2010**, 946, 70-76.

[28] M. Salanne, P. A. Madden, Molecular Physics. **2011**, 2299–2315.

[29] M. Salanne, Leonardo J. A. Siqueira, Ari P. Seitsonen, Paul A. Madden, B. Kirchner, *Faraday Discuss.* **2012**,154, 171-188.

[30] M. Salanne, Phys. Chem. Chem. Phys. **2015**, **17**, 14270-14279.

[31] M. Brüssel, M. Brehm, T. Voigt, B. Kirchner, Phys. Chem. Chem. Phys. **2011**, 13, 13617-13620.

[32] M. Brüssel, M. Brehm, A.S. Pensado, F. Malberg, M. Ramzan, A. A. Stark, B. Kirchner, *Phys. Chem. Chem. Phys.* **2012**, 14, 13204-13215.

[33] A.S. Pensado, M. Brehm, J. Thar, A.P. Seitsonen, B. Kirchner, *ChemPhysChem.* **2012**, 13(7), 1845-1853.

[34] O. O. Hollóczki, F. Malberg, T. Welton, B. Kirchner, *Phys. Chem. Chem. Phys.* **2014**, 16, 16880 - 16890.

[35] B. Kirchner, F. Malberg, D. S. Firaha, O. Hollóczki, *J. Phys.: Condens. Matter.* **2015**, 27, 463002.

[36] O. Russina, M. Macchiagodena, B. Kirchner, A. Mariani, B. Aoun, M. M. Russina, R. Caminiti, A. Triolo, *J. Non-Cryst. Solids.* **2015**, 407, 333-338.

[37] B. Brehm, H. Weber, M. Thomas, O. O. Hollóczki, B. Kirchner, *ChemPhysChem.* **2015**, 16, 3271-3277

[38] M. Campetella, M. Macchiagodena, L. Gontrani, B. Kirchner, *Mol. Phys.* **2017**. tba. DOI: 10.1080/00268976.2017.1308027

[39] S. Gabriel, J. Weiner, *Ber. Dtsch. Chem. Ges.* **1888***, 21,* 2669-2679.

[40] P. Walden, *Bull. Acad. Imper. Sci. St. Petersburg.* **1914**, 8, 405-422.

[41] S. Zahn, J. Thar, K. Barbara, *J. Chem. Phys.* **2010**, 132, 124506.

[42] D. S. Firaha, K. Barbara, *J. Chem. Eng. Data.* **2014**, 59, 3098-3104.

[43] T. Méndez-Morales, J. Carrete, Ó. Cabeza, O. Russina, A. Triolo, L. J. Gallego, L. M. Varela, *J. Phys. Chem. B.* **2014**, 118 (3), 761-770.

[44] O. Russina, R. Caminiti, T. Méndez-Morales, J. Carrete, Ó. Cabeza, Ó.; L.J. Gallego, L.M. Varela, A. Triolo, *Journal of Molecular Liquids.* **2015**, 205, 16-21.

[45] T. Méndez-Morales, J. Carrete, J. R. Rodriguez, Ó. Cabeza, L.J. Gallego, O. Russina, L.M. Varela, *Phys. Chem. Chem. Phys.* **2015**,17, 5298-5307.

[46] L.M. Varela, T. Méndez-Morales, J. Carrete, V. Gómez-González, B. Docampo-Álvarez, L.J. Gallego, Ó. Cabeza, O. Russina, *Journal of Molecular Liquids.* **2015**, 210, Part B, 178-188.

[47] R. Hayes, S. A. Bernard, S. Imberti, G. G. Warr, R. Atkin, *J. Phys. Chem. C.* **2014**, 118(36), 21215-21225.

[48] T. Ikeda, M. Boero, K. Terakura, *J. Chem. Phys.* **2007***,* 126, 034501-9.

[49] T. Ikeda, M. Boero, K. Terakura, *J. Chem. Phys.* **2015**, 143, 194510.

[50] M. J. Monteiro, F. F. C. Bazito, L. J. A. Siqueira, M. C. C. Ribeiro, R. M. Torresi, *J. Phys. Chem. B.* **2008**, 112(7), 2102-2109.

[51] O. Andreussi, N. Marzari, *J. Chem. Phys.* **2012**, 137, 044508.

[52] Z. Li, G. D. Smith, D. Bedrov, *J. Phys. Chem. B.* **2012**, 116, 12801-12809.

[53] T. Umecky, T. Takamuku, T. Matsumoto, E. Kawai, E.; M. Takagi, T. Funazukuri, *J. Phys. Chem. B.* **2013**, 117, 16219-16226.

[54] I. Skarmoutsos, V. Ponnuchamy, V. Vetere, S. Mossa, *Journal of Physical Chemistry C.* **2015,** 119, 4502-4515.

[55] O. Borodin, M. Olguin, P. Ganesh, P. R. C. Kent, J. L. Allen, W. A. Henderson, *Phys. Chem. Chem. Phys.* **2016**, 18, 164-175.

[56] K. H. Lau, J. Lu, J. Low, D. Peng, H. Wu, H. M. Albishri, D. Abd Al-Hady, L. A, Curtis, K. Amine, *Energy Technology.* **2014**, 2, 348-354.

[57] T. Li, P. Balbuena, *J. Electrochem. Soc.* **1999**, 146, 3613-3622.

[58] K. Leung, F. Soto, K. Hankins, P. B. Balbuena, K. L. Harrison, *J. Phys. Chem. C.* **2016**, *120* (12), 6302–6313.

[59] T. Katō, K. Machida, M. Oobatake, S. Hayashi, *The Journal of Chemical Physics.* **1988**, 89, 3211.

[60] T. Katō, K. Machida, M. Oobatake, S. Hayashi, *The Journal of Chemical Physics.* **1988**, 89, 7471.

[61] T. Katō, K. Machida, M. Oobatake, S. Hayashi, *The Journal of Chemical Physics.* **1990**, 92, 5506.

[62] A. K. Adya, G. W. Neilson, I. Okada, S. Okazaki, *Molecular Physics.* **1993**, 79, 1327-1350.

[63] Y. Kameda, S. Seiichi Kotani, I. Kazuhiko, Molecular Physics. **1992**, 75, 1-16.

[64] M. Iannuzzi, A. Laio, M. Parrinello, *PRL.* **2003**, 90, 238302.

[65] A. Laio, M. Parrinello, *PNAS.* **2002**, 99(20), 12562-12566.

[66] A. Barducci, M. Bonomi, M. Parrinello, *Wiley Interdisciplinary Reviews: Computational Molecular Science.* **2011**, 1, 826-843.

[67] F. Giberti, M. Salvalaglio, M. Parrinello, *IUCrJ.* **2015**, 2, 256-266.

[68] R. Hayes, S. Imberti, G. G. Warr, R. Atkin, *Phys. Chem. Chem. Phys.* **2011**, 13, 3237-3247.

[69] W. A. Henderson, P. Fylstra, H. C. De Long, P. C. Trulove, S. Parsons, *Phys. Chem. Chem. Phys.* **2012**, 14, 16041-16046.

[70] R. Hayes, S. Imberti, G. G. Warr, R. Atkin, *Angewandte Chemie International Edition.* **2013**, 52(17), 4623-4627.

[71] R. Hayes, S. Imberti, G. G. War, R. Atkin, *J. Phys. Chem. C.* **2014**, 118, 13998-14008.

[72] R. Hayes, S. Imberti, G. G. War, R. Atkin, *Chem. Rev.* **2015**, 115(13), 6357-6426.

[73] CP2K, http://www.cp2k.org/

[74] CP2K version 2.6 (Development Version), the CP2K developers group, (**2014**).

[75] S. Goedecker, M. Teter, J. Hutter, *Phys. Rev. B* **1996**, 54, 1703–1710.

[76] G. Lippert, J. Hutter, M. Parrinello, *Mol. Phys.* **1997**, 92(3), 477-487.

[77] C. Hartwigsen, S. Goedecker, J. Hutter, *Phys. Rev. B.* **1998**, 58(7), 3641-3662.

[78] M. Krack, Theor. Chem. Acc.: Theory, Comp. Modeling. **2005**, 114, 145-152.

[79] J. VandeVondele, M. Krack, F. Mohamed, M. Parrinello, T. Chassaing, J. Hutter, *Comput Phys Commun.* **2005**, 167(2),103-28.

[80] J. VandeVondele, J. Hutter, *J. Chem. Phys.* **2007**, 11, 114105.

[81] J. Hutter, M. Iannuzzi, F. Schiffmann, J. VandeVondele, *Wiley Interdisciplinary Reviews: Computational Molecular Science.* **2014**, 4(1), 15-25.

[82] M. Frigo, SG. Johnson, *PROCEEDINGS OF THE IEEE.* **2005**, **93**(2), 216-231.

[83] H. J. C. Berendsen, D. Van der Spoel, R. van Drunen, *Comput. Phys. Commun.* **1995**, 91, 43-56.

[84] E. Lindahl, B. Hess, D. van der Spoel, *J. Mol. Model.* **2001**, 7, 306-317.

[85] D. Van der Spoel, E. Lindahl, B. Hess, G. Groenhof, A.E. Mark, H. J. C Berendsen. *J. Comput. Chem.* **2005**, 26, 1701-1718.

[86] B. Hess, C. Kutzner, D. Van der Spoel, E. Lindahl, *J. Chem. Theory Comput.* **2008**, 4, 435-447.

[87] JP. Perdew, K. Burke, M. Ernzerhof, *Phys. Rev. Letters.* **1996**, 77(18), 3865-3868.

[88] S. Grimme, *J. Computational. Chemistry.* **2006**, 27, 1787-1799.

[89] S. Grimme, J. Antony, S. Ehrlich, H. Krieg, *J. Chem. Phys.* **2010**, 132(15), 154104.

[90] G. Grimme, W. Hujo, B. Kirchner, *Phys. Chem. Chem. Phys.* **2012**, 14, 4875-4883.

[91] F. Malberg, A.S. Pensado, B. Kirchner, *Phys. Chem. Chem. Phys.* **2012**, 14, 12079-12082.

[92] D. S. Firaha, M. M. Thomas, O. Hollóczki, M. M. Korth, B. Kirchner, *J. Chem. Phys.* **2016**, 145, 204502.

[93] S. Nosé, *Mol. Phys.* **1984**, 52, 255.

[94] W.G. Hoover, *Phys. Rev. A.* **1985**, 31, 1695.

[95] M. Parrinello, A. Rahman, J. Appl. Phys. **1981**, **52**, 7182.

[96] T. Ikeda, M. Boero, K. Terakura, *J. Chem. Phys.* **2007**, 127, 074503.

[97] T. Ikeda, M. Boero, *J. Chem. Phys.* **2012**, 137, 041101.

[98] I. Bakó, J. Hutter, G. Pálinkás, *J. Chem. Phys.* **2002**, 117, 9838-9843.

[99] I. Bakó, J. Hutter, G. Pálinkás, *J. Phys. Chem. A.* **2006**, 110, 2188-2194.

[100] A. W. Sekh Mahiuddin, G. Hefter, W. Kunz, B. Minofar, P. Jungwirth, J. Phys. Chem. B. **2005**, 109(50), 24108.

[101] L. Chen, G. E. Mullen, M. Le Roch, C. G. Cassity, N. Gouault, H. Y. Fadamiro, R. E. Barletta, R. A. O'Brien, R. E. Sykora, A. C. Stenson, K. N. West, H. E. Horne, J. M. Hendrich, K. R. Xiang, J. H. Davis Jr, *Angew. Chem. Int. Edt.* **2014**, 53, 11762-11765.

[102] W. Humphrey, A. Dalke, K. Schulten, *J. Molec. Graphics.* **1996**, 14, 33-38.

[103] X. Wu, F. R. Fronczek, L. G. Butler, *Inorganic Chemistry.* **1994**, 33, 1363-1365.

[104] I. Marcos-Alcalde, J. Setoain, J.I. Mendieta-Moreno, J. Mendieta, P. Gómez-Puertas, *Bioinformatics.* **2015**, 31, 3853-3855.

[105] M. Okoshi, Y. Yamada, A. Yamada, H. Nakai, *J. Electrochem. Soc.* **2013**, 160, A2160-A2165.